\documentclass[epj]{svjour}
\usepackage{graphicx}
\usepackage{txfonts}
\newcommand{\infnpd}{3}

\newcommand{\unina}{6}
\newcommand{\lngs}{1}
\newcommand{\dresden}{7}

\newcommand{\atomki}{4}

\newcommand{\unipd}{2}
\newcommand{\infnLNL}{5}

\begin{document}

\title{Underground nuclear astrophysics: why and how}

\author{
	A.\,Best \inst{\lngs} \and	
	A.\,Caciolli \inst{\unipd,\infnpd}\thanks{e-mail: caciolli@pd.infn.it} \and
	Zs.\,F\"ul\"op \inst{\atomki} \and
	Gy.\,Gy\"urky \inst{\atomki} \and
	M.\,Laubenstein \inst{\lngs} \and
	E.\,Napolitani \inst{\unipd,\infnLNL} \and
	V.\,Rigato \inst{\infnLNL} \and
	V.\,Roca \inst{\unina} \and
	T.\,Sz\"ucs \inst{\dresden} \\
	}
%
%
\institute{
	INFN, Laboratori Nazionali del Gran Sasso, Via G. Acitelli 22, 67100 Assergi (AQ), Italy 
	\and
	Dipartimento di Fisica e Astronomia, Universit\`a di Padova, and INFN, Sezione di Padova, Via Marzolo 8, Padova, Italy 
	\and
	INFN, Sezione di Padova, via Marzolo 8, 35131 Padova, Italy 
	\and 
	Institute for Nuclear Research (MTA Atomki), Debrecen, Hungary 
	\and
	INFN, Laboratori Nazionali di Legnaro, Legnaro, Italy 
	\and
	Dipartimento di  Fisica, Universit\`a di Napoli ``Federico II'', and INFN, Sezione di Napoli, Napoli, Italy 
	\and
	Helmholtz-Zentrum Dresden - Rossendorf (HZDR), Dresden, Germany 
	}

\date{\today}
%
\abstract{
The goal of nuclear astrophysics is to measure cross sections of nuclear physics reactions of interest in astrophysics. At stars temperatures, these cross sections are very low due to the suppression of the Coulomb barrier. Cosmic ray induced background can seriously limit the determination of reaction cross sections at energies relevant to astrophysical processes and experimental setups should be arranged in order to improve the signal-to-noise ratio. Placing experiments in underground sites, however, reduces this background opening the way towards ultra low cross section determination. LUNA (Laboratory for Underground Nuclear Astrophysics) was pioneer in this sense. Two accelerators were mounted at the INFN National Laboratories of Gran Sasso (LNGS) allowing to study nuclear reactions close to stellar energies. A summary of the relevant technology used, including accelerators, target production and characterisation, and background treatment is given.
} 

\authorrunning{}
\titlerunning{}

\maketitle

\section{Introduction}
\label{sec:intro}
It is well known that stars generate energy and synthesise chemical elements in thermonuclear reactions \cite{RolfsBook,IliadisBook}. All reactions induced by charged particles in a star take place in the so called Gamow peak. For the $^3$He($^4$He,$\gamma$)$^7$Be reaction, for example, which is important both for understanding the primordial $^7$Li abundance and for studying the solar neutrino problem, the Gamow peak lies at about   22.4$\pm$6.2 keV at a temperature of T$_6$\footnote{T$_6$ is the temperature expressed in T/10$^6$K} = 15, while for the Big Bang Nucleosynthesis (BBN) scenario the region of interest is about 160-380 keV.

The cross-section $\sigma(E)$ of a charged-particle induced reaction drops steeply with decreasing energy due to the Coulomb barrier in the entrance channel:
\begin{equation}\label{eq1}
\sigma(E) = \frac{S(E)}{E}\exp{(-2\pi \eta)} 
\end{equation}
where $S(E)$ is the astrophysical $S$ factor and $\eta$ is the Sommerfeld parameter with 
2$\pi \eta$ = 31.29$Z_1 Z_2 (\mu/E)^{1/2}$; $Z_1$  and $Z_2$  are the charge numbers of projectile and target nucleus, respectively, $\mu$ is the reduced mass (in amu), and $E$  is the center-of-mass energy (in keV).
In the quiescent burning of stars, the cross section is very low  at  Gamow energies and this prevents a direct measurement in a laboratory at the Earth's surface, where
the signal-to-background ratio is too small because of cosmic-ray interactions. Hence, cross-sections are measured at high energies and expressed as the astrophysical $S$  factors from eq. \ref{eq1}. The $S(E)$  factor is then used to extrapolate the data to the relevant Gamow peak. Although $S(E)$ varies  slowly with energy for the direct process, resonances and resonance tails may complicate the extrapolation, resulting in large uncertainties. Therefore, the primary goal of experimental nuclear astrophysics remains the measurement of cross-sections at energies inside the Gamow peak, or at least to approach it as closely as possible. 

The importance of the accurate determination of  cross sections arises from the fact that this parameter enters directly into the reaction rate $r = <\sigma(\nu)\cdot \nu>$, where $\sigma$ is the nuclear cross section and $\nu$ is the relative velocity of particles. 
Precise and accurate cross section data are therefore needed as inputs for stellar reaction networks.
In particular, due to the steep dependence of the cross section as a function of center-of mass energy, an accurate determination of the average energy at which a cross section is measured is of utmost importance. 
This implies that for this research field it is necessary to increase the accuracy on the measured cross sections, especially at very low energies.

\section{Experimental setups}

The realisation of the LUNA  of the INFN (Italian National Institute for Nuclear Physics) under the Gran Sasso mountain chain provided a unique opportunity to carry out in more favourable conditions the study of reactions of high astrophysical importance and to achieve results  that in a surface laboratory (or in less shielded laboratories) could not have been obtained. This goal could be reached due to the reduction of the cosmic background through the rock overburden, which diminishes the  muon flux  to $\sim$ 10$^{-8}$ cm$^{-2}$s$^{-1}$ \cite{BemmererEPJA}.
Thanks to this favourable position, LUNA was able to measure for the first time the $^3$He($^3$He, 2p)$^4$He cross-section within its solar Gamow peak \cite{Bonetti}.
Subsequently, a windowless gas target setup and a 4$\pi$-BGO summing detector have been used to study the radiative-capture reaction $^2$H(p,$\gamma$)$^3$He, also within its solar Gamow peak \cite{Casella}.
For some reactions, the underground position does not grant the possibility to measure directly at  energies of the Gamow peak. Therefore direct data at higher energies are needed and other methods can be investigated to reduce the signal to noise ratio, as for the case of Recoil Mass Separator (RMS) \cite{Gialanella}.
This apparatus allows to count few reaction products also when intense projectile beams have to be produced to overcome the small cross-sections at the interesting energies (see \cite{Gialanella} for details).

In some cases, when the reaction products are radioactive with a half life of at least  hours,   direct measurements can be followed by activation analysis on the irradiated targets \cite{Scott}, or by counting of the  reaction products  by accelerator mass spectrometry (AMS) techniques \cite{Limata25Mg}. In both cases, the irradiation can be done either in underground or in overground laboratories with an high intensity beam as that provided by LUNA.

In the case of measurements at energies higher than the Gamow peak, the value of the S factor is obtained by extrapolation  of the higher energy data. 

An interesting  case is the $^{14}$N(p,$\gamma$)$^{15}$O reaction, where to constrain the $S(0)$  extrapolation the low energy data must be combined with the high energy ones and also with experimental measurements of the resonance parameters needed in the R-matrix calculations (see \cite{marta14N,marta14N2} for details).


Low energy data taken in underground laboratory still remain a major constraint in the extrapolation. In the case of the $^3$He($^4$He,$\gamma$)$^7$Be reaction, the high precision of the low energy data provided by LUNA with two different techniques reduced the uncertainty on the $S(0)$ value of the more recent compilation  from  9\% \cite{SF1} to 3\% \cite{SF2}.

This rough outline shows that  the objectives of  Nuclear Astrophysics cannot be obtained with one technique or with one kind of instrument. The astrophysical scenarios are so wide that the proper apparatus to study a specific reaction has to be selected. In the next sections examples of different setups that can be used in underground nuclear astrophysics will be discussed.

The intensity and stability of the beam must be granted since the cross section varies steeply with the energy. Targets have to be stable over time and any possible source of beam induced background has to be minimised. Specific detection systems to follow the experimental requirements and measure the different properties of each reaction are adopted.

The main role in this context is played by the accelerators. LUNA used in its history: a 50 kV and a 400 kV machines, characterised by high-current and small energy spread, in combination with high-efficiency detection systems. The next  years will see the installation at LNGS of a new MV accelerator for widening the exploration range of the laboratory.

\section{The accelerators of LUNA}

Considering the very small cross sections and the rapid variation of their value with the energy (see eq. \ref{eq1}),  the most important characteristics of an accelerator are the beam intensity and the long term stability of the terminal high voltage (HV). 
The two accelerators used up to now at LUNA fulfil these requirements. 

The 50 kV LUNA accelerator at LNGS,  described in more details elsewhere \cite{Graife}, consisted of a duoplasmatron ion source, an extraction/acceleration system, and a double focusing 90$^\circ$ analysing magnet. The energy spread of the ion source was less than 20 eV.
The ion source, even at the lowest extraction energies, provided stable proton beams with currents of few tens of $\mu$A for periods up to 4 weeks. The HV power supply had a typical ripple of 5 $\times$ 10$^{-5}$ and a long-term stability better than 1 $\times$ 10$^{-4}$. 

The  400 kV electrostatic accelerator (High Voltage Engineering Europe, Netherlands)
was installed in 2001. It is embedded in a tank which is filled with a gas mixture N$_2$-CO$_2$  at 20 bar.  The high voltage is generated by an Inline-Cockcroft-Walton power supply (located inside the tank) capable to handle 1 mA at 400 kV.  The HV at the terminal (ion source) is filtered by a stabilisation system consisting of a RC-filter located at the output of the HV power supply and an active feedback loop based on  a chain of resistors which measures the HV at the terminal. 

In one set of experiments \cite{Formicola2003} the absolute proton energy  $E_p$ ranging between  130 keV and 400 keV was determined from the energy of the capture $\gamma$-ray transition in $^{12}$C(p,$\gamma$)$^{13}$N. The results have been checked using  (p,$\gamma$)  resonances on $^{23}$Na, $^{25}$Mg and $^{26}$Mg. These resonances were also used to measure the energy spread $\Delta E_{Beam}$  and the long-term energy stability of the proton beam.
 
The radio-frequency ion source provides ion beams of 1 mA hydrogen  (H$^+$)  and 500 $\mu$A He$^+$ over a continuous operating time of about 40 days. The ion source is mounted directly on the accelerator tube. The ions are extracted by an electrode, which is part of the accelerator tube: its voltage is thus included in the overall HV at the terminal. 
With a 45 degrees magnet  and a vertical steerer located before the magnet, the ion beam is guided and focused properly to the target station. In the energy range of 150$-$400 keV,  the proton beam current on target can reach 500 $\mu$A.
This intensity is mandatory to improve the statistics in nuclear astrophysics experiments, but it could affect the target stability due to the deterioration of solid targets (see section \ref{sect:solidtarget}) or increase the beam heating effects (see section \ref{sec:gastarget-beamheating}).
The accelerator, the experimental equipment, and the data handling are controlled by a PLC based computer, which allows for a safe operation over long periods of running time without the constant presence of an operator on site.

In 2008 a second beam line was installed to allow two experimental setups to be mounted simultaneously.

\section{Nuclear Astrophysics Underground}
\label{sec:underground}

LUNA has focused its work on quiescent hydrogen burning, with relevant temperatures in a range from 20 to 100 MK \cite{BrogginiReview,CostantiniReview}. The corresponding Gamow peak lies at energies from 10 to hundreds of keV. Measuring at these energies is challenging because of  the drastic reduction of the cross section due to the Coulomb barrier. The experimental statistics can be improved by increasing the beam intensity, the target density and the detection efficiency, but it can also be achieved by reducing the two components of background: laboratory and beam induced.

\subsection{Laboratory Background}\label{sec:bck}

The laboratory background comes either from the signal produced by cosmic rays or from the one produced by radionuclides present in the experimental environment.
In  gamma spectra each of these two components dominate a specific energy range.
In particular, the $\gamma$-ray spectrum at energies up to 2.6 MeV is dominated by the signal of radionuclides, especially $^{40}$K and the decay chains of $^{238}$U and $^{232}$Th.

A careful selection procedure of setup materials is required to minimise this background and the setup is usually shielded with passive materials (i.e. copper and lead). 
Active shielding can also be  applied, increasing the complexity of the acquisition electronics.

\begin{figure}[htb]
    \includegraphics[width=\columnwidth]{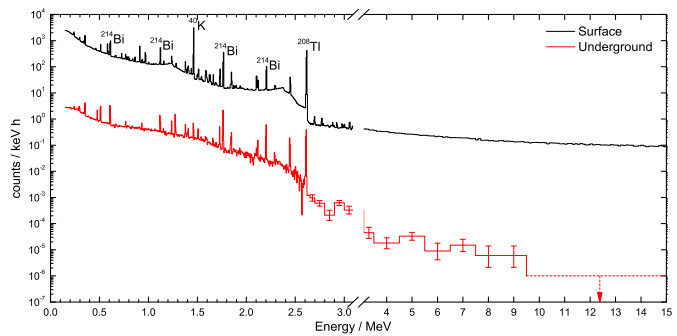}
    \caption{Environmental background acquired with a HPGe detector on surface laboratory (black) and inside the LNGS (red). At LNGS the HPGe detector was included in a shielding made of lead and copper as discussed in \cite{cav14}.}
    \label{fig:bck}
\end{figure}

At energies above 3 MeV, the background in the $\gamma$ spectrum  is basically due to  cosmic radiation.
At sea level, this radiation consists of 70\% muons, 30\% electrons, and $<$1\%  protons and neutrons. Muons are the most penetrating component and they produce background in counting facilities.
They lose energy by ionisation and can contribute to the detector background in different ways: i) by losing energy while traversing the detector itself; ii) by producing energetic electrons which induce secondary electrons and $\gamma$ radiation; iii) by producing interactions in materials surrounding the detectors followed by X-rays, $\gamma$ rays, and neutron emission.
In addition, neutrons produced by muon spallation induce  radioactivity  in the experimental environment.
Such background can be overcome by performing experiments in a deep underground laboratory, where the muon flux is greatly reduced. For example in the INFN National Laboratories of Gran Sasso  the muon flux above 1 TeV is reduced by 6 orders of magnitude thanks to its depth of 3800 m.w.e. (meters of water equivalent)  \cite{BemmererEPJA,TsucsEPJA}. This is clearly shown  in figure \ref{fig:bck} where two spectra acquired with the same detector are compared. The spectrum in black is acquired on surface while the spectrum in red is acquired inside the LNGS with the detector shielded with copper and lead (the details of the setup are discussed in \cite{cav14}). In the $\gamma$ spectrum, above 4 MeV  the reduction is due to the underground position, while below 3 MeV the effect of the shielding is clearly evident. The main background lines are labeled in the black spectrum.

The integral background counting rate from 40 to 2700 keV for High Purity Germanium (HPGe) detectors, often used in underground laboratories, as a function of the shielding depth levels off at a certain depth, even though the muon part of the cosmic rays is still decreasing exponentially with depth. 
This is a clear indication that other background sources become more important, i.e. intrinsic natural radioactivity, neutrons from spontaneous fission and ($\alpha$,n)-reactions, and residual cosmogenic activation from the above-ground production of the detector. Of course, also  radon gas becomes important and has to be removed carefully from the environment around the detector \cite{Caciolli2009}. In addition, a prerequisite for obtaining ultra-low background is, of course, that the detector is radio pure. 

It has to be noted that the background reduction from cosmic radiation has a relevant effect also for the $\gamma$-energy region below 3 MeV. This is evident if we compare shallow underground laboratories with the results achieved at the LNGS with a fully shielded setup \cite{BrogginiReview}.

\subsection{Neutron background}

Besides the effect on $\gamma$ spectroscopy, locating an experiment deep underground  strongly influences also backgrounds in charged particle spectroscopy \cite{BrunoEPJA} and neutron detection.
The latter is an important aspect considering the great interest in neutron sources for s-process in nuclear astrophysics.

The astrophysical s process is responsible for the production of about half of the elements heavier than iron \cite{Kappeler1999}. Its two main neutron
sources are  the $^{13}$C$(\alpha$,n$)^{16}$O and $^{22}$Ne$(\alpha$,n$)^{25}$Mg reactions \cite{Aliotta}. The standard approach to measure this kind of reactions consists in using a moderating neutron detector, e.g., \cite{Falahat2013}. The thermalisation process implies the loss of information on the initial energy of the measured
neutrons. 

Therefore, to be able to measure at the very low energies required for the astrophysical scenarios of the s process, the  neutron background needs to be suppressed as much as
possible. For the  reaction $^{22}$Ne$(\alpha$,n$)^{25}$Mg \cite{Jager} the energy region of interest lies around 200 - 300 keV, below the lowest already measured data point. The natural neutron background remains the limiting factor towards lower energies. To reach the stellar energy region another two or three orders of magnitude of background reduction
are necessary. This can be done by going deep underground.

The neutron flux on the surface of the earth is dominated by cosmic-ray induced neutrons. Generally, it is of the order of $10^{-3}$ cm$^{-2}$s$^{-1}$. By going deep underground one can reduce the cosmic ray
flux by 6 orders of magnitude. Then, the radiogenic component becomes the main neutron background, two to four orders of magnitude higher than the cosmogenic neutron flux. As already discussed in section \ref{sec:bck}, the underground neutron
sources are mostly spontaneous fission of $^{238}$U in the cavity walls and $(\alpha,n)$ reactions induced by $\alpha$-particles from the natural radioactivity of the underground environment.
\begin{figure}[htb]
    \includegraphics[width=\columnwidth]{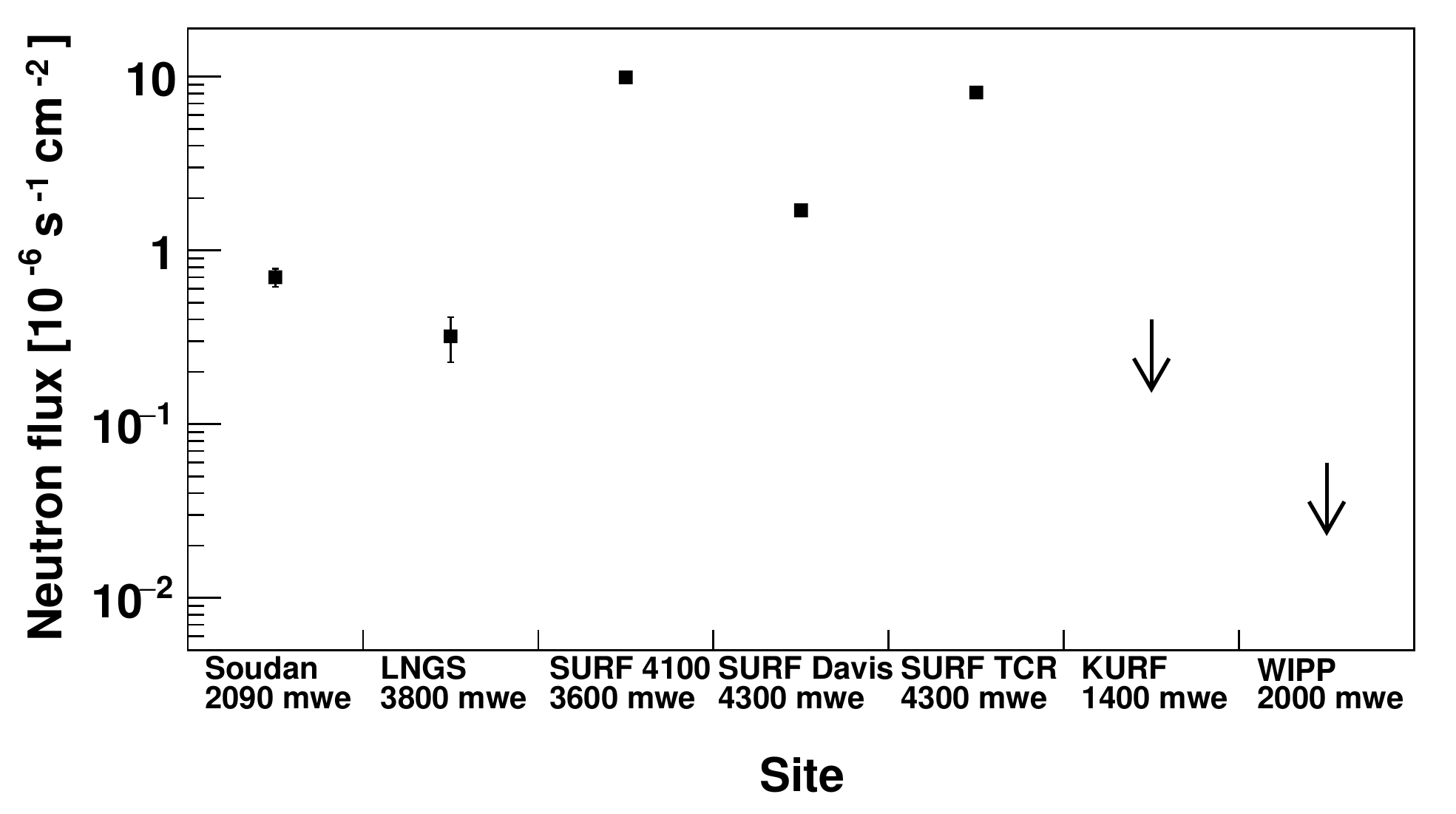}
    \caption{Thermal neutron flux at various deep underground laboratories.}
    \label{fig:neutron-fluxes}
\end{figure}

The exact value of the neutron flux is very sensitive to the local
environment, like the composition of the shotcrete and the surrounding rocks and on the radon concentration in the vicinity of the detector. The thermal component of the neutron background
comprises the majority of the total flux and has therefore the most significant impact on the sensitivity of experiments, especially when using neutron counter detectors.
Various measurements of the thermal neutron background at the LNGS have been done; none of them agree with each other, possibly due to a variation in the background depending on the location in the laboratory or due to unknown systematic uncertainties. The reported values are: (2.05$\pm$0.06)$\cdot$10$^{-6}$cm$^{-2}$s$^{-1}$ \cite{Rindi}, (1.08$\pm$0.02)$\cdot 10^{-6}$cm$^{-2}$s$^{-1}$ \cite{Belli}, and (5.4$\pm$1.3)$\cdot$10$^{-7}$cm$^{-2}$s$^{-1}$ \cite{Debicki}. A recent measurement shown in figure \ref{fig:neutron-fluxes}  agrees with the latter value \cite{Best}.
Figure \ref{fig:neutron-fluxes} shows a comparison of the thermal neutron fluxes in different deep underground laboratories measured with $^{3}$He counters. It can be seen that even in the same laboratories the flux can vary by up to an order of magnitude, making efficient air ventilation and local shielding
an important component in the planning of new facilities. At LNGS the thermal flux is below $10^{-6}$ cm$^{-2}$s$^{-1}$,   the background suppression is at a level
so far not reached by surface experiments.

\subsection{Beam Induced Background}

Environmental background is a critical issue for direct measurements in nuclear astrophysics but in experiments with ion beams, impurities in the apparatus and in the targets can induce reactions that  give rise to background (mainly $\gamma$-ray and neutron background).

Such beam induced background is independent of the underground depth and must be dealt with by reducing the inventory of materials hit by the ion beam and taking appropriate precautions to eliminate worrisome components.

\begin{figure}
\centering
\includegraphics[width=\columnwidth]{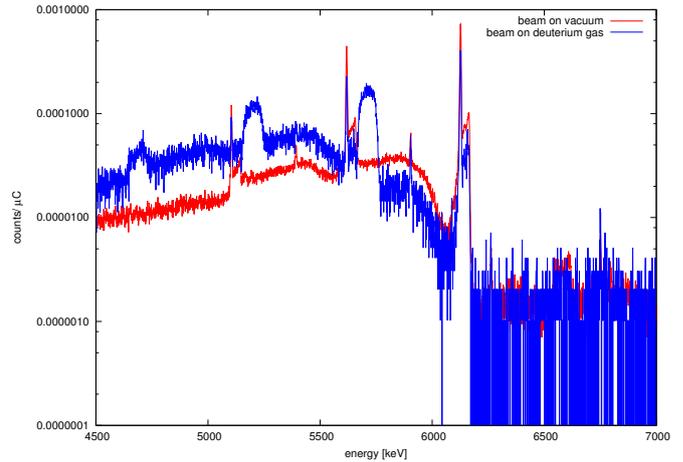}
\caption{\label{fig:fluorine} Comparison of two spectra acquired at the same proton beam energy using as target vacuum  (in red) and  deuterium (in blue). The peaks due to the p+d reaction are clearly visible in the blue spectrum, but they are not present in the blue one where the peaks due to the $^{19}$F(p,$\alpha \gamma$)$^{16}$O reaction still remain. This suggests that the fluorine is deposited on the surfaces of the experimental setup.}
\end{figure}

Main contaminants are low Z elements like deuterium, boron, carbon, oxygen, and fluorine. Contributions from elements with higher Z are usually negligible at astrophysical energies. 
Fluorine is one of the most common contaminants that can be found and the $^{19}$F(p,$\alpha \gamma$)$^{16}$O reaction has a high cross section at energies below 400 keV (the one of interest for LUNA400kV accelerator). In particular, there is a resonance at $E_p$ = 340 keV with a large width that produces many structures in the 5-8 MeV energy region of the $\gamma$-ray spectrum. In figure \ref{fig:fluorine} a spectrum acquired with deuterium gas as target for the study of the $^2$H(p,$\gamma$)$^3$He reaction is shown in blue. The peaks due to the reaction of interest and the corresponding structures due to the beam induced background on $^{19}$F are visible in the spectrum. For comparison a spectrum acquired without gas in the target chamber is also shown (in red):  only the peaks from the  $^{19}$F(p,$\alpha \gamma$)$^{16}$O reaction are still visible. 

\begin{figure}
\centering
\includegraphics[width=\columnwidth]{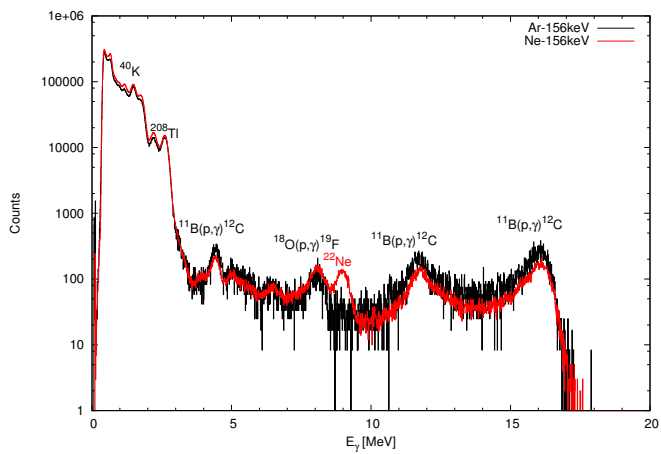}
\caption{\label{fig:boron} Comparison of two spectra acquired at the same proton beam energy using as target argon gas  (in black) and  neon gas (in red). The peaks due to the reactions on boron and oxygen  are clearly visible, as discussed in the text.}
\end{figure}

Boron and oxygen are other common sources of contamination. Especially boron is frequently present in the experimental surfaces and it is difficult to remove it. The (p,$\gamma$) reaction on boron-11 mainly emits three $\gamma$ rays: around 4.4 MeV, 12 MeV and 16 MeV. In figure \ref{fig:boron}, the boron signal is clearly visible both in the spectrum with argon as target and in the one with neon (enriched in $^{22}$Ne) filling the scattering chamber. Another peak is visible in both spectra, due to the $^{18}$O(p,$ \gamma$)$^{19}$F reaction. Subtracting the two spectra, it is possible to obtain a clear peak at the energy of interest of the $^{22}$Ne(p,$\gamma$)$^{23}$Na reaction (labelled in figure \ref{fig:boron}).

When the effort of reducing contaminants that produce beam induced background in reaction studies is not successful, careful study of the reactions involved in the beam induced background can help in understanding their behaviour and their contribution to the acquired signal \cite{AndersEPJA}. Many of these reactions are already deeply studied in literature and this helps in managing their contribution in data analysis  comparing spectra acquired with blank samples \cite{Bemmerer2009} and/or using Monte Carlo simulations \cite{Strieder25Mg}. 

\section{Gas Target}
\label{sec:gastarget}

In nuclear astrophysics, reaction cross section measurements very often need the use of a gas target system. This is especially true in the case of reactions relevant for hydrogen and helium burning processes, where the cross section measurement on the isotopes of gaseous elements like hydrogen (deuteron), helium, nitrogen, oxygen, neon, etc. are necessary. In some cases, experiments can also be performed with solid state targets using a suitable compound, like TiN or Ta$_2$O$_5$ in the case of nitrogen or oxygen, respectively \cite{for04,sco12}. These targets, however, have their disadvantages concerning, e.g. the limited target stability and the uncertain stoichiometry. Moreover, no chemical compound can be formed from the noble gases, reducing the stability of solid targets for these elements. Therefore, gas target systems are extensively used in nuclear astrophysics experiments and represent an important aspect of underground experiments as well.

Besides being unavoidable in some cases, gas targets also have some clear advantages over solid state targets. The target thickness can be easily regulated by changing the gas pressure and the number of target atoms can be determined rather precisely using the pressure and temperature measurements. No target degradation occurs, which is a clear advantage in nuclear astrophysics where typically high beam intensities are encountered in order to measure low cross sections.

Gas targets can be categorised into two distinct groups: windowless gas targets and thin window gas cells. For the latter, the beam enters the gas target through a thin metal window which separates the target setup from the accelerator \cite{bor12}. In underground nuclear astrophysics experiments, low energy beams are typical not exceeding  the few hundred keV range. In this energy range, even the thinnest possible entrance window would cause too high energy loss and large energy straggling increasing substantially the uncertainty of the measurements. Thin window gas cells are therefore usually avoided in low energy nuclear astrophysics. 

Windowless gas targets can be of extended or jet type. Quasi point-like gas targets can be formed with a jet. In this case, a supersonic gas stream is let to the target chamber through a small nozzle where it is hit by the particle beam. Since jet targets were never used at LUNA, this section concentrates on extended gas targets. Details about jet targets can be found e.g. in \cite{kon12}.

\subsection{Properties and characterisation of an extended gas target}
\label{sec:gastarget-properties}

An extended gas target consists of a finite volume of gas, a gas filling system, pumping stages and some auxiliary equipment for, e.g. gas purification or beam intensity measurement. The physical length of the gas volume is typically between a few centimetres up to several tens of centimeters. The volume is confined either by two small apertures, or only one aperture towards the accelerator. In the latter case the beam stop is placed inside the gas volume.

\begin{figure*}
\centering
\includegraphics[width=0.8\textwidth]{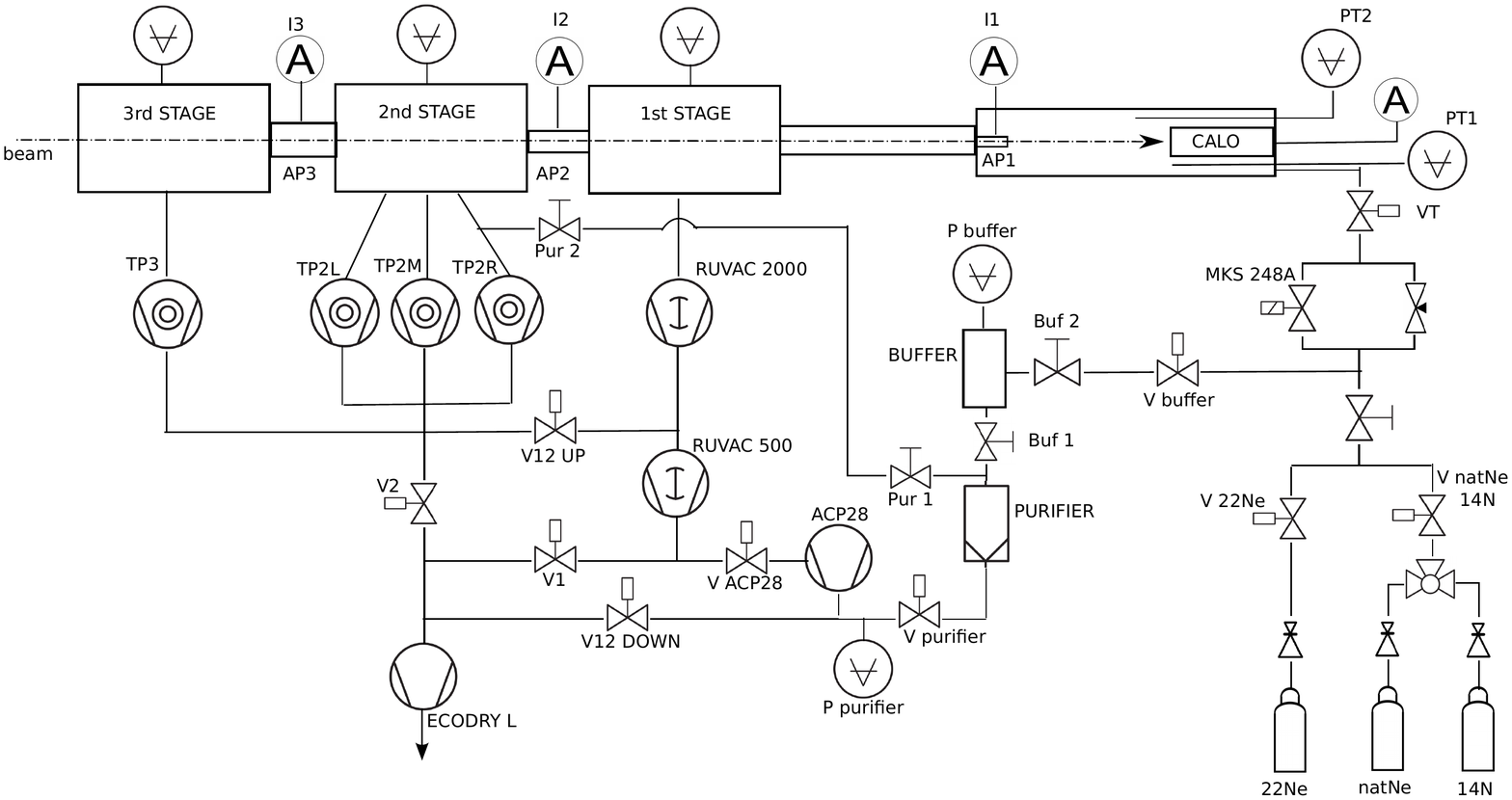}
\caption{\label{fig:gastarget} Block diagram of a typical extended gas target pumping system \cite{cos06,cav14}.}
\end{figure*}

Extended gas targets operate typically at pressures in the mbar range. The accelerator on the other hand requires high vacuum, i.e. a pressure of several orders of magnitude lower  than in the gas target. As there is no window separating the gas target from the rest of the beam line, the necessary pressure drop is realized by a series of apertures and pumping stages. In the first pumping stage the heavy load of target gas is pumped away by high  speed pumps, like a roots blower, resulting in a pressure drop of roughly 3 orders of magnitude. The subsequent pumping stages are not pumping such a high amount of gas and therefore turbomolecular pumps can be used to reach the required vacuum of the accelerator.

Since gas is  lost through the apertures, the target gas must be continuously supplied in order to keep a constant pressure in the target. This is achieved by gas inlet valves regulated by the pressure sensors of the target system. In the case of expensive gases, like isotopically enriched ones, it is necessary to recirculate the gas lost from the target. In such a case, the outlets of the pumping stages are fed back to the gas inlet system. In order to keep the chemical purity of the gas, a purifier may be necessary. Such a purifier can ideally be used in the case of a noble gas target, as the air contamination of the gas is efficiently absorbed by the chemical getter of the purifier, while the noble gas is let through. Figure \ref{fig:gastarget} shows the block diagram of a typical extended gas target system. Such a system was used by LUNA in the case of several experiments, e.g. $^3$He($\alpha$,$\gamma$)$^7$Be \cite{bem06}, $^2$H($\alpha$,$\gamma$)$^6$Li \cite{and14} and $^{22}$Ne(p,$\gamma$)$^{23}$Na \cite{cav15}.

For a cross section measurement, one of the parameters which must be known is the number of target atoms per unit area. In the case of an extended gas target this number is obtained from the physical length of the chamber and the pressure and temperature of the gas. If the cross section is measured, e.g. by the detection of the prompt $\gamma$-radiation emitted in the studied capture reaction, then an effective target length is defined which takes into account those parts of the target from where the $\gamma$-radiation can reach the detector.  

Any pressure variation along the target chamber must also be taken into account. Therefore, the pressure profile of the target must be determined by measuring the pressure of the gas at several positions, along the beam line. The pressure profile outside the actual gas target, in the first pumping stage, must also be known as a non-negligible part of the total target thickness can be located there. Especially if a high temperature beam calorimeter is used (see below), the temperature profile of the target gas must also be measured precisely in order to obtain an accurate gas density and hence target thickness \cite{cav14}.
Of course the pressure measurement alone  does not give any information about the composition of the gas. Any gas impurity can falsify the determined number of target atoms and can also cause unwanted background. Gas impurities (most typically nitrogen leaking in from the atmosphere) can be studied  by sampling the gas occasionally and measuring its composition with a quadrupole mass spectrometer. Another possibility is the measurement of the elastic scattering of the beam particles in the gas. Details about such a method can be found in \cite{mar06}.

\subsection{Beam current measurement}
\label{sec:gastarget-current}

The other important input quantity of a cross section measurement is the number of projectiles impinging on the target. In the case of a solid state target setup, this quantity is usually determined through charge integration.  In the case of a gas target setup, however, the beam particles passing through the gas undergo charge exchange reactions with the target atoms and secondary electrons are also produced. Therefore, the beam particles arrive  at the beamstop with highly uncertain charge state together with the created secondary electrons. The simple charge integration method is thus not applicable.

Measurement of the elastic scattering of the projectile on the gas target nuclei can in principle give information about the product of the number of target atoms and projectiles. This method is often used in the case of a jet gas target. For extended targets, on the other hand, this method is not ideal as it gives information only about a small part of the target length and therefore introduces high uncertainties. 

The usual solution is the application of beam calorimetry. Here, instead of measuring the charge carried by the beam to the beam stop, the deposited power is measured. If the beam energy is known, the beam intensity and therefore the number of projectiles can be obtained. A precise way of measuring the beam power is the application of a constant gradient beam calorimeter. In such a device, a constant temperature difference is sustained between the thermally coupled cold and hot side of a calorimeter by a power resistor. If the beam is heating the beam stop, the resistor is able to keep the temperature difference with a reduced power. The beam intensity can be calculated from the difference between the power measured with and without beam \cite{Casella2002}. 

In order to precisely determine the beam intensity, the beam calorimeter must be calibrated. This can be done using an evacuated gas target and charge integration. Without gas, the beam intensity can be measured with the conventional charge integration method. If the beam power is measured simultaneously with the calorimeter, then the latter one can be calibrated. An important condition for such a calibration is that the chamber around the beam stop forms a good Faraday cup for a reliable charge measurement.

\subsection{Beam heating effect}
\label{sec:gastarget-beamheating}

If an intense particle beam passes through a gas target, the power of the beam is in part deposited in the gas itself causing a local heating effect. As a result of the local temperature rise, the gas density and hence the effective target thickness decrease. This effect cannot be observed with the pressure sensors as they are located outside the beam path. 

There are different ways to measure the beam heating effect. The total energy loss of the beam in the gas target depends on the gas density along the beam path and therefore on the beam heating effect. If the energy of the beam at a given point along the beam path can be measured precisely, then the reduced energy loss and thus the beam heating can be measured. A nuclear reaction exhibiting a narrow resonance can be used for such a measurement. The excitation function around the resonance is measured with different beam intensities and gas pressures, and the beam heating can be derived from the position of the resonance edge \cite{gor80}. 

Another possibility is offered by the detection of elastically scattered beam particles on the gas as the yield of the scattering changes linearly with the gas density. The elastic scattering yield can be measured at different gas pressures and the measurement can be extrapolated down to zero pressure where no beam heating occurs. The measurement can be carried out at various beam energies and intensities and, if the flexibility of the scattering setup allows, at different positions along the beam line. Such an extensive study was carried out at LUNA in relation to the $^3$He($\alpha,\gamma$)$^7$Be experiment and a beam heating effect of up to 10\,\% was found with beam intensities of typically 300\,$\mu$A and gas pressure of up to 1.3\,mbar \cite{mar06}. 


\section{Solid Target}
\label{sect:solidtarget}

The second beamline of the LUNA-400kV is devoted to reaction studies with solid targets. 
The advantages of a gas target setup have been explained in section \ref{sec:gastarget}. 
The solid target approach can be chosen when gas target handling is complicated or too  expensive, and most importantly in cases where the isotope to be studied is not of a gaseous element, and/or does not form any reasonable gas compound. 
In addition, a solid target can be considered a point-like source in experiments involving $\gamma$ emission and this simplifies the experimental needs to study  angular distributions and/or correlations.

Due to the reduced dimension of the targets, the scattering chambers can fit easily the request of the detection setup and it is usually easier to put the detectors in close geometry (which is important in low statistics measurements).
As already discussed briefly in section \ref{sec:gastarget}, the disadvantages of solid targets setups are due to the difficulties in characterization and stability of the target samples. Therefore a careful study of the targets  with several ion beam analysis techniques is needed.

\subsection{The LUNA solid target beamline}

The beam passes through the 0$^\circ$ port of the first dipole magnet  and then it is bent by 45$^\circ$ in                                a second dipole magnet. 
The beam is collimated  using two apertures between analysing magnet and the target.
An electric triplet is also placed between the analysis magnets to improve the beam focusing.

Solid targets are susceptible to the deposition of contaminants on their surface.
While in the gas target setup new gas is constantly introduced into the scattering chamber, solid targets should be kept free from contaminant depositions on their  surface.
Contaminants decrease the beam energy and cause also beam straggling.
This is a critical problem in nuclear astrophysics measurements where the cross section changes drastically with the beam energy. 
Therefore the vacuum system should be optimised in order to have pressures in the range of 10$^{-7}$ mbar.
This goal can be achieved also by introducing a cold trap close to the target position. 
In the LUNA solid target setup, a 20 cm long copper tube is placed at 2 mm  from the target. 
The tube is cooled with LN$_2$ in order to catch residual contaminant particles in the last part of the beamline. Thanks to this system, no build-up of any contaminant has been observed during long time irradiation in LUNA experiments \cite{DiLeva,Caciolli15N}.

\subsection{Target production and analysis}

\begin{figure}
\includegraphics[angle=0,width=\columnwidth]{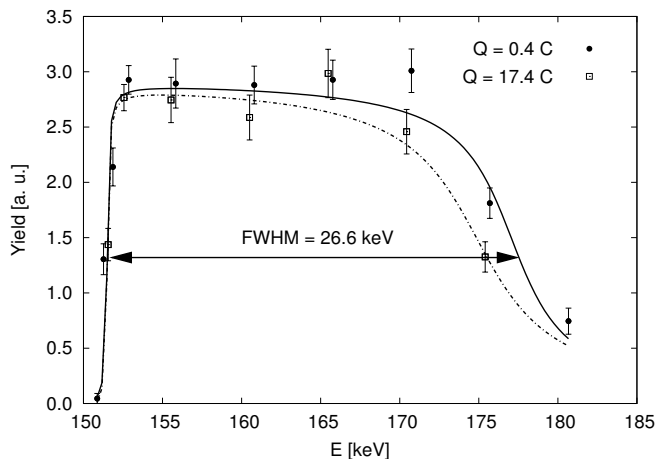}
\caption{\label{fig:targetscans} Thick-target yield profile for the 151~keV resonance in
$^{18}$O(p,$\gamma$)$^{19}$F  for two values of accumulated charge on the same Ta$_2$O$_5$ target. The error bars are statistical only. The fitted value for the target thickness is reported in the figure. The dashed line is a fit to the
thick-target yield profile obtained after a total accumulated
charge on target of 17.4 C: a reduction in thickness is visible.}
\end{figure}

There are several techniques used in solid target production. The most common ones  are implantation, reactive sputtering, oxidation, and evaporation. 
Reactive sputtering TiN targets were used at LUNA to study the $^{14}$N(p,$\gamma$)$^{15}$O and $^{15}$N(p,$\gamma$)$^{16}$O reactions, thanks to their great stability under beam bombardment and their very precise stoichiometry. 
Experiments at LUNA are characterised by high beam intensities in order to maximise the statistics and therefore the targets are critically stressed. For the production of  Ta$_2$O$_5$ targets, a dedicated setup was installed at the LNGS \cite{CaciolliTa2O5}. Those targets were shown to keep the same characteristics in stoichiometry and isotopic ratio after 20 C of accumulated charge.
As already outlined,  in gas target setups the gas is constantly refurbished in the scattering chamber by means of the circulating system.  Solid targets should be constantly monitored in order to keep under control their characteristics. 
Narrow resonances with well known  strength can be used for this purpose. The scan of the integrated yield of the resonance contains many  parameters of the target composition such as the  thickness, $\Delta E$, and the number of active atoms. The yield for a narrow resonance can be written as:
\begin{equation}\label{eq:inf_target}
Y = \frac{\lambdabar ^2}{2}\frac{\omega \gamma}{\epsilon_r} 
\end{equation}
if the energy loss in the target thickness is much higher than the resonance width ($\Delta E >> \Gamma_{tot}$) \cite{IliadisBook}. In eq. \ref{eq:inf_target}, $\lambdabar$ is the De Broglie wavelength, $\omega \gamma$ is the resonance strength and $\epsilon_r$ the effective stopping power.
 The latter is the stopping power weighted for the reactive ions in the target. 
Repeating this scan in regular time intervals (see Figure \ref{fig:targetscans}) allows to monitor targets consumption or contaminant deposition on the target surface (i.e. observing a shift in the front edge of the resonance scan). 

Contaminants are one of the most problematic issues of solid targets. 
Contaminants can be introduced during  target preparation, be present in the backings (e. g. fluorine in tantalum backings) or they can be deposited during irradiation from residual particles in the beamline vacuum. 
Impurities in the targets are removed by carefully polishing the  target backings and the target preparation setups and with proper target handling  procedures. Most common contaminants are fluorine, carbon, oxygen, and deuterium.

\begin{figure}
\includegraphics[angle=0,width=\columnwidth]{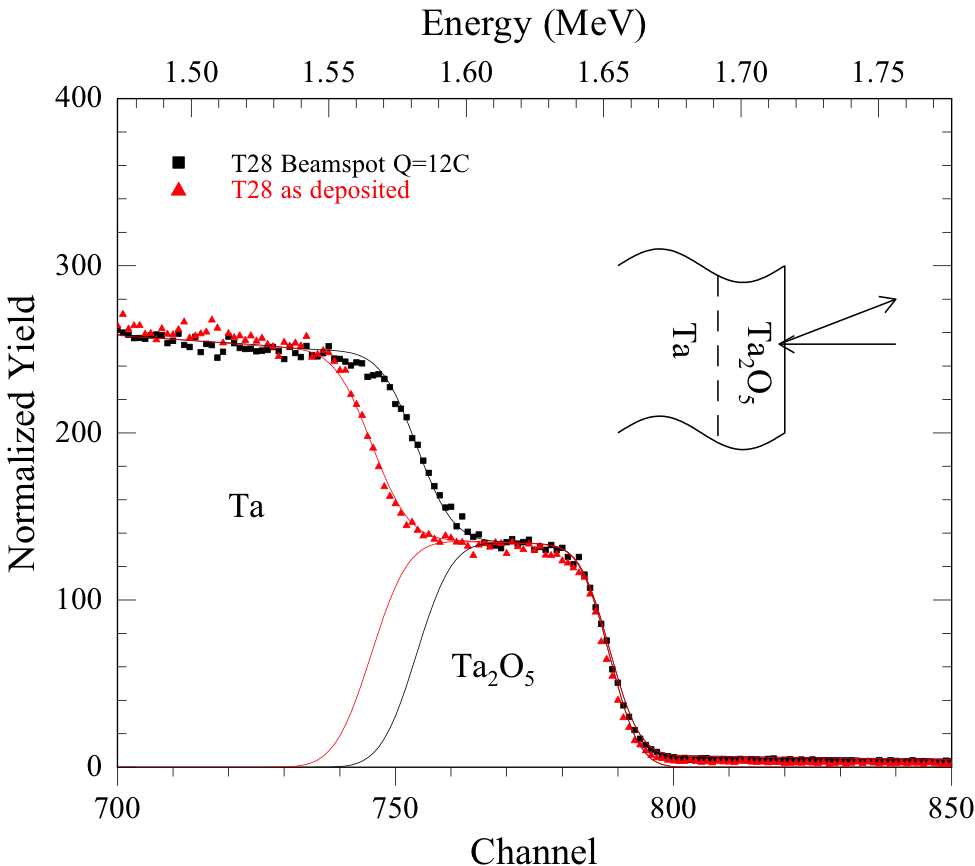}
\caption{\label{fig:RBS} RBS spectra measured with 2 MeV $\alpha$ beam on a Ta$_2$O$_5$ sample. The spectra are acquired after different accumulated charges on the sample. The signal from the oxygen is not visible, while the layer of  tantalum oxide is clearly separated from the Ta backing as shown in the simulation (lines). After  12~C  accumulated charge (black spectrum) the reduction of thickness is clearly visible without any change of the stoichiometry Ta:O with respect to the original target (red).}
\end{figure}

If targets cannot be checked online during the irradiation or if it is not possible to extrapolate all information from a single technique, offline analysis has to be used. Typical techniques  are Rutherford Backscattering (RBS), Secondary Ionised Mass Spectrometry (SIMS), Elastic Recoil Detection Analysis (ERDA), and Nuclear Reaction Analysis (NRA). 
The contents of the TiN targets with nitrogen enriched in $^{15}$N were investigated by a resonance scan of the 429.5~keV resonance of the $^{15}$N(p,$\alpha \gamma$)$^{12}$C reaction at the HZDR Tandetron accelerator \cite{MartaHZDR} and with the High Z ERDA technique at the Q3D magnet \cite{Bergmeier} to understand the isotopic ratio achieved during the target preparation. 
In offline analysis, the samples are investigated in the region irradiated by the beam during the experiment and in a region outside the beamspot and comparison are also made with not irradiated samples to describe in the best way the sample behaviour during high intensity beam irradiation.
In Figure \ref{fig:RBS} two spectra acquired with $\alpha$ backscattering technique on Ta$_2$O$_5$ samples are shown. These measurements were done at the AN2000 accelerator of the INFN National Laboratories of Legnaro using an $\alpha$ beam with a beam size of 1~mm$^2$ and $\alpha$ energy of 1.8 and 2.0~MeV.
The elastically scattered $\alpha$ particles were detected by a silicon detector at 160$^\circ$ with respect to the beam axis. 
The sample is Ta$_2$O$_5$ layer on Ta backing with oxygen enriched in $^{18}$O at 90\%. The black spectrum is acquired after an irradiation of 12~C at the LUNA-400kV and it is possible to derive the reduction of the Ta$_2$O$_5$ layer thickness, while the stoichiometry, which can be deduced from the height  of the plateau at channels 770-790, remains stable also after long irradiation. RBS and NRA are useful to get information on different elements and isotopes and their chemical composition in  different layers. Therefore with a unique non destructive measurement, many pieces of information (i.e. thickness, isotopic and chemical compositions, contaminants) can be achieved reducing costs and time needs.

SIMS  was used for the Ta$_2$O$_5$ targets in order to determine the isotopic enrichment in $^{17}$O and $^{18}$O \cite{CaciolliTa2O5}.  SIMS is a powerful technique \cite{SIMS} to investigate also the abundance profile along the target thickness. Differently from RBS, it is not affected by the problem of high Z backings that  overwhelms the RBS spectrum. 

\begin{figure}
\includegraphics[angle=0,width=\columnwidth]{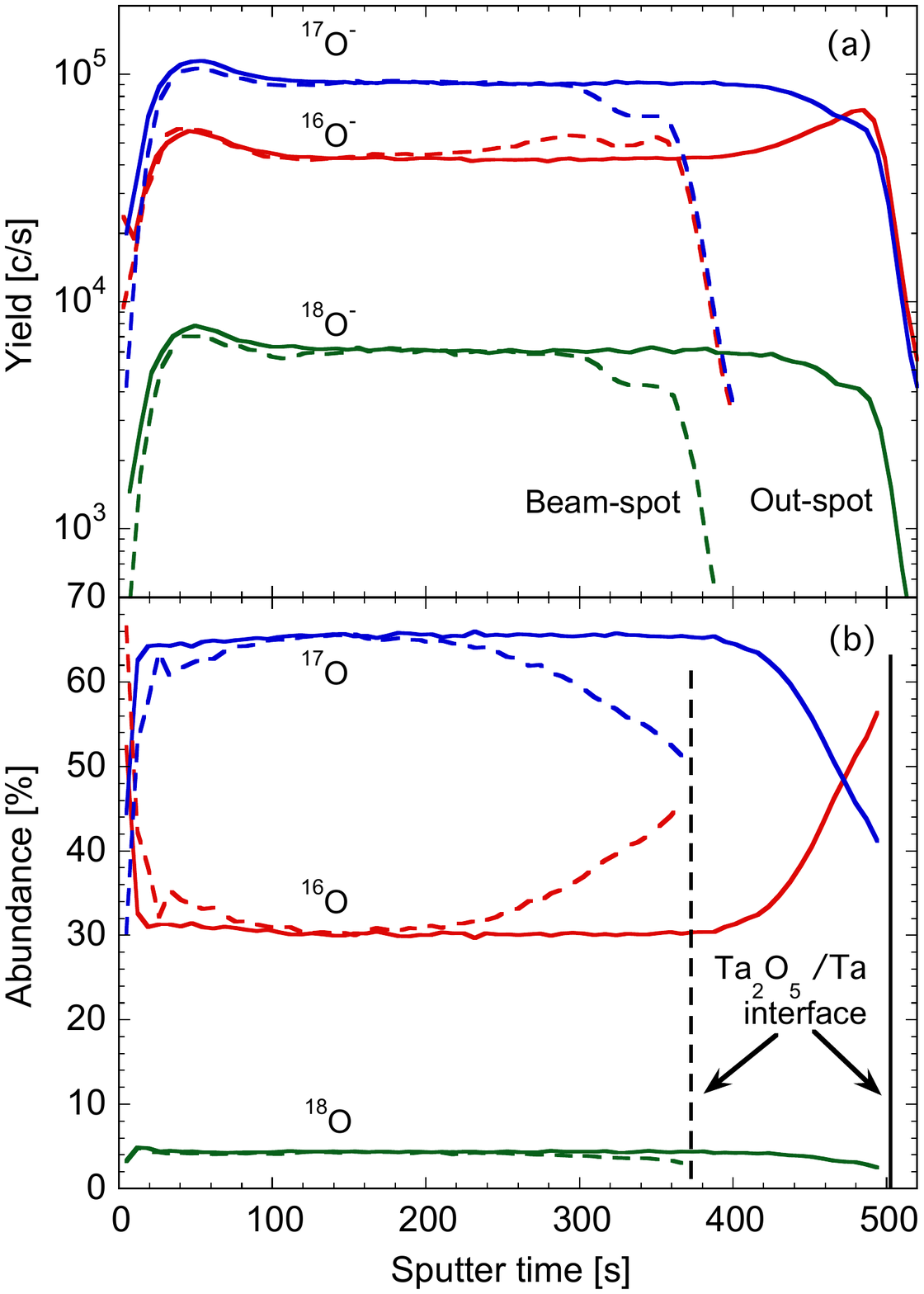}
\caption{\label{fig:SIMS} (Colour online) (a) Depth yield profiles of secondary $^{16}$O$^-$, $^{17}$O$^-$ and $^{18}$O$^-$ ions as a function of the erosion time, as obtained by Secondary Ion Mass Spectrometry analyses of a Ta$_2$O$_5$/Ta sample enriched in $^{17}$O. (b) $^{16}$O, $^{17}$O and $^{18}$O isotope abundance depth profiles calculated from the yield profiles reported in panel (a). The depth of the Ta$_2$O$_5$/Ta interfaces are also indicated with vertical lines. Measurements were taken in the region of the target exposed to the proton beam at LUNA (beam-spot, dashed lines) and in peripheral regions not exposed to the proton beam (out-spot, continuous lines).}
\end{figure}

In Figure \ref{fig:SIMS}, an example of a SIMS study on the Ta$_2$O$_5$ irradiated at LUNA is shown. Two regions of the target were selected to study its properties before and after the usage at  LUNA-400kV. During SIMS analysis, the sample surface is continuously sputtered away by a focused and rastered primary ion beam. Secondary isotope ions that are emitted from the sample surface are selected with a high resolution mass spectrometer and eventually collected. The result is a yield profile of the selected isotopes as a function of the erosion time.  Depth profiles of isotopic abundances of $^{16}$O, $^{17}$O and $^{18}$O in LUNA isotopically enriched Ta$_2$O$_5$/Ta layers were measured by SIMS at the Department of Physics and Astronomy of the University of Padua (Italy), using a CAMECA IMS-4f spectrometer.
A 14.5~keV, 3~nA Cs$^+$ primary beam is rastered over a crater area of 100 x 100 $\mu$m$^2$. The secondary ions $^{16}$O$^-$, $^{17}$O$^-$, and $^{18}$O$^-$ are collected during sputtering only from a central region of 30 $\mu$m diameter in order to avoid crater-edge effects. High mass resolution (M/$\Delta$M = 5000) is used in order to eliminate mass interferences of $^{17}$O$^-$ and $^{18}$O$^-$ with the molecular ions $^{16}$O$^1$H$^-$ and $^{17}$O$^1$H$^-$, respectively. Measurements of a pure Si sample were compared to the natural Si isotopic abundance as a check for systematic errors.
Fig. \ref{fig:SIMS}a shows the $^{16}$O$^-$, $^{17}$O$^-$ and $^{18}$O$^-$ yield profiles as a function of the erosion time collected both in out-spot and beam-spot regions on a representative target enriched with $^{17}$O. Fig. \ref{fig:SIMS}b reports the corresponding isotope ratio profiles extracted from the SIMS profiles of Fig. \ref{fig:SIMS}a. Data clearly show a broad central thickness with flat abundances. 
Outside the plateau, yields may be affected by several phenomena, such as transients of the sputtering process due to the surface and  presence of contaminants at the surface and at the Ta$_2$O$_5$/Ta interface. An additional $^{16}$O contamination is always observed in beam-spot measurement, most probably due to beam-induced O migration both from the surface and from the Ta$_2$O$_5$/Ta interface. However, its effect on the total thickness remains usually negligible. In addition, beam-spot measurement shows a thickness reduction, indicated by the drop in the yields occurring at earlier sputter times with respect to the out-spot sample, in agreement with results from RBS and NRA.

Combining the information from all these IBA techniques it is possible to reach a precision of few percent on the number of target atoms, which is the goal of nuclear astrophysics studies.

\section{Conclusions}

LUNA started its activity almost 25 years ago with the goal of exploring the fascinating domain of nuclear astrophysics at very low energy.
During its activity, LUNA has proven that direct measurements in nuclear astrophysics  benefit from going in an underground environment, like the National Laboratories of Gran Sasso. For the first
time, the important reactions that are responsible for hydrogen burning in the Sun have been
studied down to the relevant stellar energies. 
In recent years, many other similar projects have been started:  new accelerators will be installed  in underground laboratories, e.g., LUNA-MV  in Europe \cite{Aliotta}, CASPAR in the US \cite{CASPAR}, and JUNA in China \cite{JUNA}. The new LUNA-MV machine will start operating in 2018 and will be focused on helium burning reactions \cite{Aliotta}.

\section*{Acknowledgments}
Financial support by INFN, FAI, DFG (BE 4100-2/1) and OTKA (K101328) is gratefully acknowledged.

%
\bibliographystyle{epjc}

\end{document}